%% file: tpdi_long.tex
\pdfoutput=1
\documentclass[10pt,pdftex,a4paper,notitlepage]{iopart_jf}

\usepackage[papersize={21cm,29.7cm}]{geometry}
\newcommand{\titlestr}{Electron correlation in two-photon double ionization of helium from attosecond to FEL pulses}

\usepackage{placeins}
\usepackage{amsmath,amssymb}

\usepackage{graphicx,color}
\usepackage[numbers,sort&compress,comma]{natbib}
\usepackage{hyperref}
\usepackage{hypernat}

\definecolor{rltred}{rgb}{0.75,0,0}
\definecolor{rltgreen}{rgb}{0,0.5,0}
\definecolor{rltblue}{rgb}{0,0,0.75}
\hypersetup{colorlinks,%
hypertexnames=true,%
pdfauthor={Johannes Feist, Renate Pazourek, Stefan Nagele, Emil Persson, Barry I. Schneider, Lee A. Collins, Joachim Burgd\"orfer},%
pdftitle={\titlestr},%
pdfpagemode=UseNone,%
bookmarksopen=true,%
bookmarksnumbered=true,%
pdfhighlight=/I,%
linkcolor=rltred,%
citecolor=rltgreen,%
linktocpage=true}

\input{custom_defs}

\begin{document}
\twocolumn[\input{frontmatter}]
\input{introduction}
\input{method}
\input{cross_secs}
\input{pci}
\input{shakeup}
\input{summary}
\FloatBarrier

\input{biblio}
\end{document}

%% file: custom_defs.tex
\newcommand{\E}{\mathcal{E}}

\newcommand{\ie}{i.e., }
\newcommand{\Schro}{Schr\"o\-din\-ger }
\newcommand{\eg}{e.g.\@ }
\newcommand{\cf}{cf.~}

\newcommand{\abs}[1]{\left|#1\right|}
\newcommand{\ket}[1]{|#1\rangle}

\renewcommand{\equationautorefname}{Eq.}

\newcommand{\cvec}[1]{\mathbf{#1}}
\newcommand{\unitv}[1]{\mathbf{\hat{#1}}}
\newcommand{\eqcomma}{\,,}

\newcommand{\dd}{\mathrm{d}}
\newcommand{\dt}{\dd t}
\newcommand{\dE}{\dd E}
\newcommand{\hw}{\hbar\omega}

\newcommand{\submax}{\mathrm{max}}
\newcommand{\Lmax}{L_\submax}
\newcommand{\lonemax}{l_{1,\submax}}
\newcommand{\ltwomax}{l_{2,\submax}}

\newcommand{\ev}{\,\mathrm{eV}}
\newcommand{\au}{\,\mathrm{a.u.}}
\newcommand{\Wcm}{\,\mathrm{W}/\mathrm{cm}^2}
\newcommand{\as}{\,\mathrm{as}}
\newcommand{\fs}{\,\mathrm{fs}}
\newcommand{\He}{\mathrm{He}}
\newcommand{\Hep}{\He^+}
\newcommand{\kone}{{\cvec{k}_1}}
\newcommand{\ktwo}{{\cvec{k}_2}}

\newcommand{\Tp}{T_{\mathrm{p}}}
\newcommand{\Tmax}{T_{\mathrm{max}}}
\newcommand{\Eeqs}{E_\mathrm{eq}}
\newcommand{\Emax}{E_\mathrm{max}}
\newcommand{\Esi}{E_\mathrm{SI}}
\newcommand{\rsi}{r_\mathrm{SI}}
\newcommand{\ti}{t^{(i)}_c}
\newcommand{\Ti}{T^{(i)}_c}
\newcommand{\tii}{t^{(ii)}_c}
\newcommand{\Tii}{T^{(ii)}_c}
\newcommand{\tioii}{t^{(i,ii)}_c}
\newcommand{\Tioii}{T^{(i,ii)}_c}
\newcommand{\Tc}{T_c}
\newcommand{\DE}{\Delta E}
\newcommand{\limTpinf}{\lim_{\Tp\to\infty}}
\newcommand{\PI}{P^{I}}
\newcommand{\PDI}{P^{DI}}
\newcommand{\PDIs}{\PDI_\mathrm{seq}}
\newcommand{\PDIns}{\PDI_\mathrm{nonseq}}
\newcommand{\PDIrel}{\PDI_\mathrm{rel}}
\newcommand{\PDInonres}{\PDI_\mathrm{nonres}}
\newcommand{\Pasym}{\mathcal{A}}

%% file: frontmatter.tex
\title[Electron correlation in TPDI of He from attosecond to FEL pulses]{\titlestr}

\author{J.~Feist$^1$, R.~Pazourek$^1$, S.~Nagele$^1$, E.~Persson$^1$, B.~I.~Schneider$^{2,3}$, L.~A.~Collins$^4$, and J.~Burgd\"orfer$^1$}
\address{$^1$ Institute for Theoretical Physics, 
              Vienna University of Technology, 1040 Vienna, Austria, EU}
\address{$^2$ Physics Division, 
              National Science Foundation, Arlington, Virginia 22230, USA}
\address{$^3$ Electron and Atomic Physics Division, 
              National Institute of Standards and Technology, Gaithersburg, Maryland 20899, USA}
\address{$^4$ Theoretical Division, T-4,
              Los Alamos National Laboratory, Los Alamos, New Mexico 87545, USA}
\ead{johannes.feist@tuwien.ac.at}
\date{\today}

\begin{abstract}
We investigate the role of electron correlation in the two-photon double ionization of
helium for ultrashort XUV pulses with durations ranging from a hundred attoseconds to 
a few femtoseconds. We perform time-dependent 
\emph{ab initio} calculations for pulses with mean frequencies in the so-called ``sequential''
regime ($\hw>54.4\ev$).
Electron correlation induced by the time correlation between emission events manifests itself in the angular 
distribution of the ejected electrons, which strongly depends on the 
energy sharing between them. We show that for ultrashort pulses two-photon double
ionization probabilities scale non-uniformly with pulse duration depending on the energy sharing
between the electrons. Most interestingly we find evidence for an interference between
direct (``nonsequential'') and indirect (``sequential'') double photo-ionization with intermediate
shake-up states, the strength of which is controlled by the pulse duration. This observation may
provide a route toward measuring the pulse duration of FEL pulses.

\end{abstract}

%% file: introduction.tex
\section{Introduction}\label{sec:intro}

The role of electron correlation is of central interest
in our understanding of atoms, molecules and solids. The recent progress in
the development of ultrashort and intense light sources
\cite{HenKieSpi2001,SanBenCal2006,GouSchHof2008,FLASH2007,NabHasTak2005,%
DroZepGop2006,NauNeeSok2004,SerYakSer2007,ZhaLytPop2007,NomHorTza2008} provides unprecedented
opportunities to study the effects of correlation not only in stationary states,
but also in transient states (\ie resonances), and even to actively induce dynamical
correlations \cite{FeiNagPaz2008b}.

The helium atom is the simplest atomic system where electron-electron interactions
can be studied, with its double ionization being the prototype reaction for a three-body
Coulomb breakup. While computationally challenging, the full dynamics of the helium atom 
can still be accurately simulated in \emph{ab initio} calculations \cite{ParSmyTay1998}.
With the advent of intense XUV pulses, the focus has shifted from single-photon double 
ionization \cite{ByrJoa1967,ProSha1993,BraDoeCoc1998,BriSch2000,MalSelKaz2000} and intense-IR
laser ionization by rescattering (\cite{LeiGroEng2000a,StaRuiSch2007,RudJesErg2007} and references therein)
to multiphoton ionization. Two-photon double ionization (TPDI) has recently received considerable attention,
both in the so-called ``nonsequential'' or ``direct'' regime ($39.5\ev < \hw < 54.4\ev$), where the electrons
necessarily have to share energy via electron-electron interaction to achieve double ionization
\cite{NikLam2001,ColPin2002,FenHar2003,HuCoCo2005,FouLagEdaPir2006,IvaKhe2007,NikLam2007,%
ProManMar2007,HorMccRes2008,FeiNagPaz2008,AntFouPir2008,GuaBarSch2008,HasTakNabIsh2005,SorWelBob2007},
and in the ``sequential'' regime ($\hw>54.4\ev$), where electron-electron interaction is not a priori
necessary \cite{LauBac2003,PirBauLau2003,IshMid2005,BarWanBur2006,HorMorRes2007,FouAntBac2008,FeiNagPaz2008b}.

In a previous paper \cite{FeiNagPaz2008b}, we investigated the role of energy and angular correlations 
in the shortest pulses available today, where the distinction between ``sequential'' and ``nonsequential'' 
becomes obsolete.
In this contribution, we explore the dependence of two-photon double ionization (TPDI) on the pulse duration $\Tp$ ranging 
from $\sim\!\!100$ attoseconds (the duration of the shortest pulses produced by high-harmonic generation \cite{GouSchHof2008})
to a few femtoseconds (the expected duration of FEL pulse ``bursts''). 
$\Tp$ can be used as a control knob to change from a ``direct'' to an ``indirect'' process.
In \autoref{sec:tpdi}, we discuss the behavior of the one-electron ionization rate $\PDI(E)/\Tp$, which displays 
non-uniform scaling with $\Tp$. In \autoref{sec:pci}, we investigate the angular correlations, with a focus on longer 
pulses, which reveal the detailed dynamics of the TPDI process.
In \autoref{sec:shakeup}, we show that for energies above the threshold associated with
shake-up ionization of the He atom, interferences between sequential and nonsequential
contributions can be observed, the strength of which can be varied by changing the pulse duration.
One consequence is that from the size and shape of these 
Fano-like resonances, the pulse duration of XUV pulses might be deduced. 
All this information is encoded in the final joint momentum distribution 
$\PDI(\cvec{k}_1,\cvec{k}_2) \equiv \PDI(E_1,E_2,\Omega_1,\Omega_2)$, which is
experimentally accessible in kinematically complete COLTRIMS measurements \cite{UllMosDor2003}. 
In this contribution, we focus on integrated quantities, which are more readily accessible
because of better statistics. Unless otherwise stated, atomic units are used.

%% file: method.tex
\section{Method}\label{sec:method}

Our theoretical approach (described in more detail in \cite{FeiNagPaz2008}) 
is based on a direct solution of the time-dependent \Schro equation (TDSE)
by the time-dependent close-coupling (TDCC) scheme \cite{ColPin2002,LauBac2003,HuCoCo2005,PinRobLoc2007}. 
The TDSE is solved in its full dimensionality including all inter-particle interactions. 
The laser field is linearly polarized and treated in dipole approximation. The interaction operator is 
implemented in both length and velocity gauge, such that gauge independence can be explicitly checked.
In the TDCC scheme the angular part of the wave function is expanded in coupled spherical harmonics. For the discretization of the radial functions, we employ a finite element discrete variable representation (FEDVR)~\cite{ResMcc2000,MccHorRes2001,Schneider05,SchColHu2006}. A local DVR basis within each finite element leads to a diagonal representation of all potential energy matrices.
The sparse structure of the kinetic energy matrices enables efficient parallelization, giving us the possibility to employ pulses with comparably long durations (up to a few femtoseconds) in our simulations.
For the temporal propagation of the wave function, we employ the short iterative Lanczos (SIL) method \cite{ParkLight86,SmyParTay1998,Lefo90} with adaptive time-step control. 

Dynamical information is obtained by projecting the wave packet onto products of Coulomb 
continuum states. As these independent-particle Coulomb wave functions are not solutions 
of the full Hamiltonian, projection errors are, in principle, inevitable. However,
since we are able to propagate the wavepacket for long times after the conclusion of the pulse, 
errors in the asymptotic momentum distribution can be reduced to the one-percent level by 
delaying the time of projection until the two electrons are sufficiently far apart from 
each other \cite{FeiNagPaz2008}.

Most of the results presented were obtained at a mean photon energy of $\hw=70\ev$, which would correspond to 
the sequential regime for long pulses. We choose the vector potential to be of the form
$\cvec{A}(t)=\unitv{z}A_0\sin^2(\pi t/(2\Tp))\sin(\omega t)$ for $0<t<2\Tp$. The duration 
$\Tp$ corresponds to the FWHM of the sine-squared envelope function.
The peak intensity was chosen as $I_0=10^{12}\Wcm$ to ensure that ground state
depletion and three-or more photon effects are negligible. 
In order to reach convergence of the 
angular distribution, single electron angular momenta up to values of $\lonemax=\ltwomax=10$ were used.
The highest total angular momentum included in 
the time propagation was $\Lmax=3$. For extracting the final probability distributions, only the two-photon
channels $L=0$ and $L=2$ were taken into account. The radial grid was composed of FEDVR elements of $4\au$ 
with order $11$, with an extension up to $r_\submax=800\au$ for the longest pulses.
All presented quantities were tested for numerical convergence and gauge independence.

%% file: cross_secs.tex
\section{Pulse length dependence of TPDI}\label{sec:tpdi}

The nature of the two-photon double
ionization (TPDI) process depends strongly on the photon energy. 
In order to doubly ionize the helium atom (ground state energy
$E_0\approx -79\ev$), each photon must have an energy
of at least $\hw = -E_0/2 \approx 39.5\ev$. 
For $39.5\ev<\hw<54.4\ev$, a single photon does
not provide sufficient energy to ionize the $\Hep$ ion. Thus, TPDI can
only occur if the two electrons exchange energy during the ionization
process. In a temporal picture, this implies that the ``first'',
already ejected, electron still has to be close to the nucleus when the
second photon is absorbed, \ie both photons have to be absorbed
quasi-simultaneously (or nonsequentially).
For photon energies larger than the ground state energy of the $\Hep$ ion $(\hw>54.4\ev)$,
an independent-particle picture is applicable for long pulses: each electron absorbs one
photon and electron-electron interaction is a priori not required for
double ionization to occur. The first electron is released from the He atom
with an energy of $E_1=\hw-I_1$, while the second electron is
released from the $\Hep$ ion with an energy of $E_2=\hw-I_2$. Here,
$I_1\approx24.6\ev$ ($I_2\approx54.4\ev$) is the first (second)
ionization potential of helium. For shake-up satellites the partitioning
of ionization potentials is different ($I_{2}'=I_2/n^2$), and so are the
peak positions $E_{1,2}'$, but the overall picture of sequential and independent
photoionization events remains unchanged.

For ultrashort pulses of a few
hundred attoseconds, the notion of sequentiality loses its meaning. 
The breakdown of the independent-particle picture and
strong coupling between the outgoing electrons is in that case not
imposed by the necessity of energy-sharing but is enforced by the ultrashort
time correlation between the two photoemission events occurring within $\Tp$.
Electron-electron interaction therefore plays a decisive role in the
correlated final momentum distribution. In particular,
the electrons are preferably emitted in a back-to-back configuration at
approximately equal energy sharing, corresponding to a Wannier ridge 
configuration \cite{FeiNagPaz2008b}.

\begin{figure*}[tbp]
  \centering
  \includegraphics[width=\linewidth]{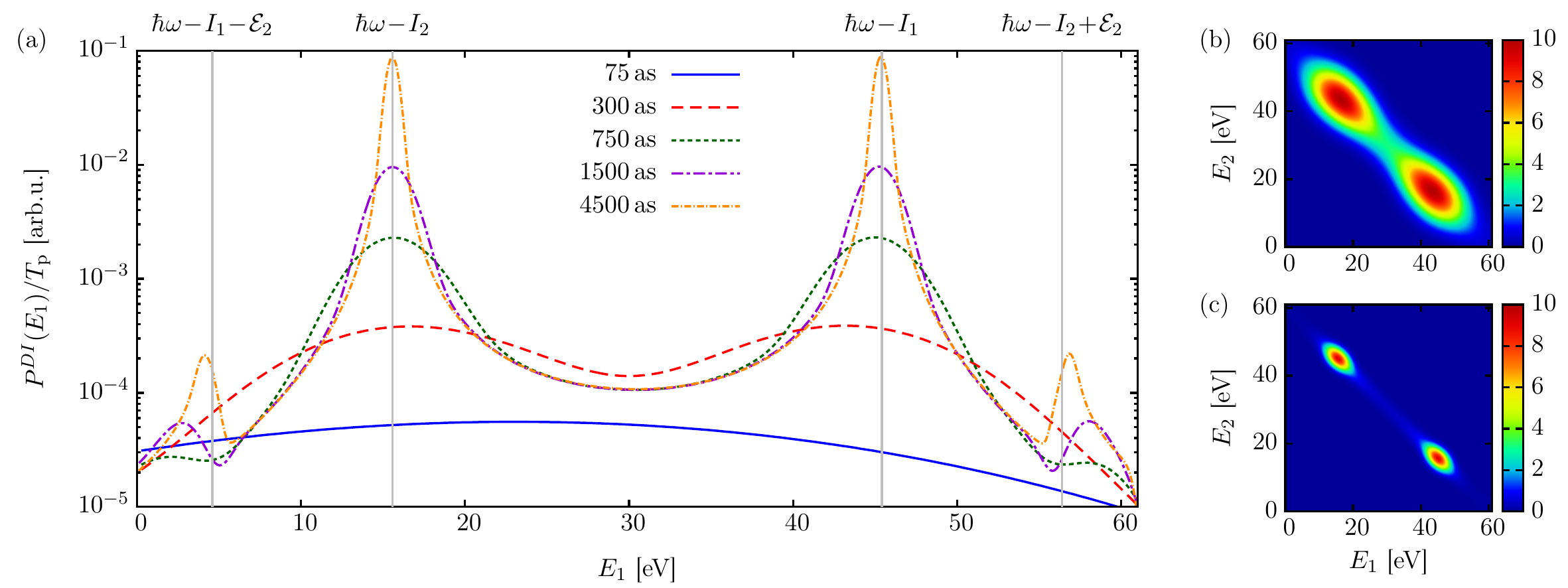}
  \caption{(a) Double ionization (DI) rate $\PDI(E)/\Tp$ (\ie DI probability divided by the pulse duration) 
  for TPDI by an XUV pulse at $\hw=70\ev$ with different pulse durations $\Tp$.
  For sufficient pulse duration, the DI rate converges to a stable value except near
  the peaks of the sequential process. (b) and (c) show the two-electron energy spectrum
  $\PDI(E_1,E_2)$ for (b) $\Tp=300\as$ and (c) $\Tp=750\as$.}
  \label{fig:pdi_dur_scan}
\end{figure*}

A key indicator for sequential TPDI is that for sufficiently low intensities 
(when ground state depletion is negligible), the total
yield scales with $\PDIs\propto\int_{-\infty}^\infty \int_t^\infty I(t)I(t')\dt'\dt 
\propto \Tp^2$, where $\Tp$ is the
duration of the laser pulse \cite{FeiNagPaz2008,IshMid2005}.
This is an immediate consequence of two independent subsequent emission processes,
the probability for each of which increases linearly with $\Tp$, such that $\PDIs\sim\!(\PI)^2\sim\Tp^2$. 
Equivalently, for each of the two processes a well-defined transition rate $W=\limTpinf\PI/\Tp$
exists. This implies that the total rate $\PDIs/\Tp$ of the two-step process grows linearly with
$\Tp$ in the limit of long pulses.
By contrast, the nonsequential or direct double ionization probability $\PDIns$ scales linearly with
$\Tp$ and a converged transition rate exists in the limit $W=\limTpinf\PDIns/\Tp$.

\begin{figure*}[tbp]
  \centering
  \includegraphics[width=\linewidth]{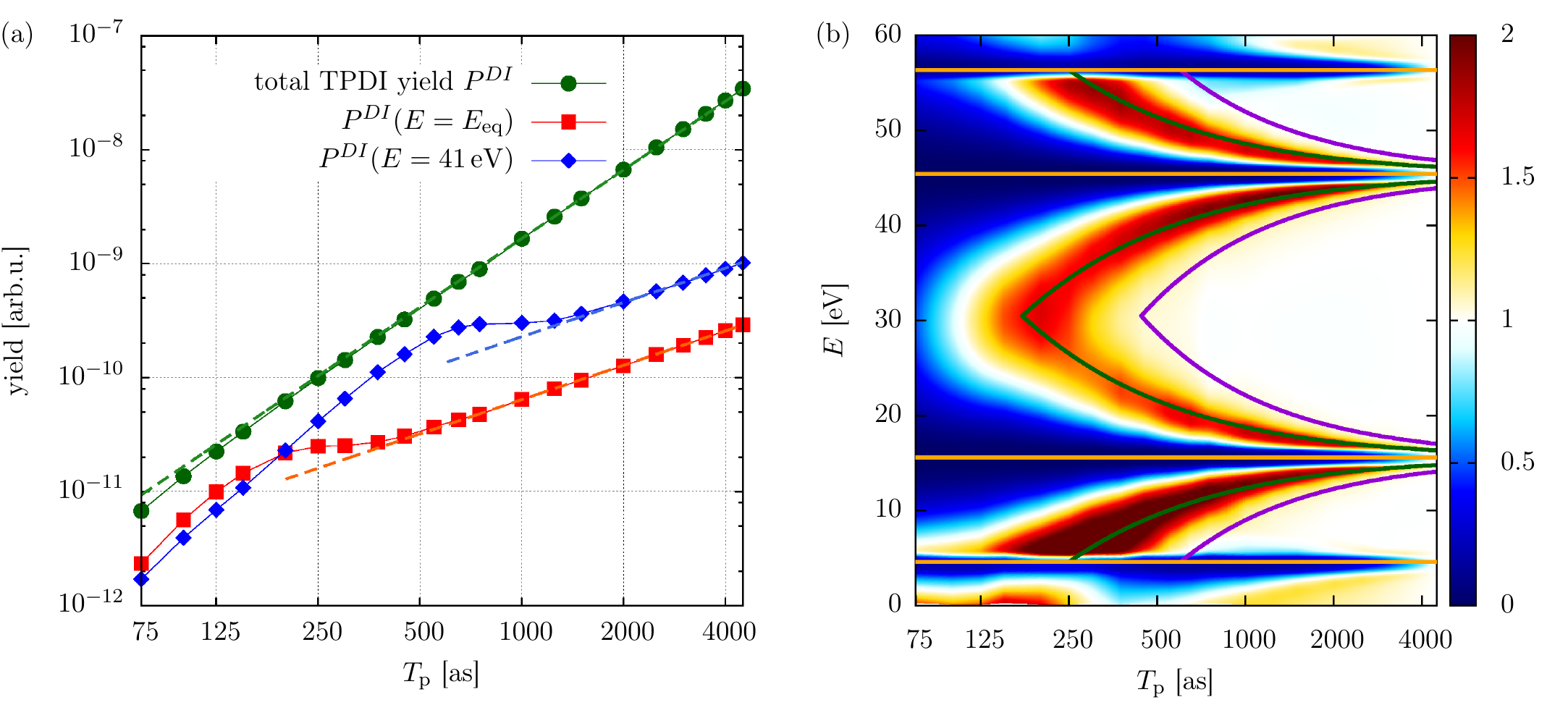}
  \caption{(a) Scaling of two-photon double ionization yields with pulse
  duration $\Tp$ (FWHM of the $\sin^2$ XUV pulses) at $\hw=70\ev$. 
  The green points are the total ionization
  yield $\PDI$, the red squares give the differential yield at equal
  energy sharing $\PDI(E=\Eeqs)$, with $\Eeqs=(2\hw+E_0)/2$, and the
  blue diamonds give the differential yield at $E=41\ev$. 
  The dashed lines show fits to quadratic and linear scaling with $\Tp$ for
  the total and singly differential yield.
  (b) Contour plot of $\PDI_{rel}(E,\Tp)$. A value of $1$ for $\PDI_{rel}$ (white in the color scale used here)
  marks the region where linear scaling of the singly differential yield with pulse duration $\Tp$ is observed.
  The orange lines indicate the positions of the peaks from the sequential process. 
  The violet and green lines indicate the pulse durations $\Ti$ and $\Tii$ after which linear scaling
  of the yield with $\Tp$ is expected due to Fourier broadening of the sequential peak 
  and because of the maximum time delay between the photon absorptions (see text).}
  \label{fig:yield_tp_scaling}
\end{figure*}

For ultrashort pulses, the scaling of the ionization yield with $\Tp$ varies between $\Tp$ and $\Tp^2$ highlighting
the non-uniform convergence over different regions of the electron emission spectrum and
the breakdown of the distinction between direct and indirect processes. \autoref{fig:pdi_dur_scan}a
illustrates the dependence of the energy differential electron emission probability 
(projection of the joint energy distribution \autoref{fig:pdi_dur_scan}b,c, onto the $E_1$ (or $E_2$) axis)
for different pulse durations, divided by $\Tp$, $\dd W/\dd E=\PDI(E)/\Tp$. This quantity
converges to a duration-independent cross section value (apart from constant factors) except in the regions near
$E=\hw-I_1$ and $E=\hw-I_2$, \ie those values of the energy where the sequential 
process is allowed \cite{HorMorRes2007}.
The peak areas grow linearly with $\Tp$ indicative of an overall quadratic scaling characteristic 
for the sequential process (\cf\autoref{fig:yield_tp_scaling}a). If one divides the yield contained in the peak areas by $\Tp^2$, the 
result is just proportional to the product of the single ionization
cross sections for one-photon absorption from the He ground state and
one-photon absorption from the $\Hep$ ground state.

The region within which the linear scaling prevails is determined by the pulse duration for two different reasons:

(i) Due to Fourier broadening, the photon energy is not well defined for a 
finite pulse, limiting the energy resolution.
Thus, if the broadened sequential peak overlaps with the final energy
of interest, the long-pulse limit $\PDI(E)\propto\Tp$ can not be observed.

(ii) There is an intrinsic maximum time 
delay between ionization events that can lead to a specific combination of final energies of the
ejected electrons. When the second electron is ionized at a time when the first electron is 
already far from the nucleus, the electrons cannot exchange a sufficient amount of energy.
For each final state, there is 
a maximum delay $\tii$ between ionization events that can lead to that specific energy sharing.
This implies that the pulse has to be considerably 
longer than this maximal delay in order to resolve all contributions to a specific final state.

In order to estimate the size of effect (ii), we employ a simple classical model:
the first electron is emitted with energy $\Esi=\hw-I_1$. In order to reach a specific final state 
with energies $(E_1,E_2)$, the liberated electron has to gain or lose the energy $\DE=\min(\abs{\Esi-E_1},\abs{\Esi-E_2})$
by interacting with the second electron.
Therefore, the first electron can be at most a distance $\rsi(\tii)=1/\DE$ from the core at the 
moment of the second photon absorption. This leads to a critical time
\begin{equation}\label{eq:linscal_pred}
\tii  =  \frac{2\sqrt{\alpha(\alpha+1)} -
    \ln\left(2\alpha + 2\sqrt{\alpha(\alpha+1)} + 1\right)}{(2\Esi)^{3/2}} \eqcomma
\end{equation}
with $\alpha=\Esi/\DE$. 

Likewise, the spectral width of the pulse gives a corresponding time $\ti=1/\DE$. 
Linear scaling should be observed for pulse durations $\Tc$ much longer than $\tioii$.
Setting $\Tioii \approx 10\,\tioii$ leads to good agreement with the full numerical simulation 
(\autoref{fig:yield_tp_scaling}b). Moreover, both criteria give similar results thereby
precluding a clear distinction between them. \autoref{fig:yield_tp_scaling}b displays
the estimates $\Tioii$ and the fraction of double ionization probability that scales linear 
with $\Tp$ as a function of emission energy and pulse duration,
\begin{equation}\label{eq:pdi_rel}
\PDIrel(E,\Tp)= \frac{\PDI(E,\Tp)}{\PDI(E,\Tmax)}\frac{\Tmax}{\Tp} \eqcomma
\end{equation}
where $\Tmax=4.5\fs$ is the longest pulse we used. $\PDIrel$ takes on the value one when
the double ionization probability at energy $E$ shows linear scaling with pulse duration.
We note that the estimate of effect (ii) could be validated in a time-independent perturbation theory calculation.
The latter does not show Fourier broadening but introduces an effective cutoff for the interaction time
$\tii$ because of the limited box size.

For long enough pulses, there is an additional interesting feature at
energies $E_1=\hw-I_1-\E_2$ and $E_2=\hw-I_2+\E_2$, with
$\E_n=2-2/n^2$ the excitation energy to the $n$-th excited state in
$\Hep$. At these energies, sequential ionization via the excited ionic (shake-up) state
$\ket{nl}$ is allowed. We discuss this is in more detail in \autoref{sec:shakeup}.

The non-uniform scaling with $\Tp$ described here should occur for any 
photon energy where the sequential process is allowed.
This is confirmed by calculations at $\hw=91\ev$, shown in \autoref{fig:91ev_pdi_dur_scan}.
At these higher photon energies, the ionized electrons obtain higher momenta,
such that larger box sizes are required in the simulation for the same pulse duration.
Therefore, the maximum duration of the laser pulse was restricted to $\Tp=1.5\fs$.

%% file: pci.tex
\section{Angular correlations}\label{sec:pci}
Additional information on the dynamics of the two ionized
electrons can be extracted from the angular correlations in the TPDI
process. To that end, we introduce
the forward-backward asymmetry distribution $\Pasym(E_1,E_2)$,
obtained by fixing the ejection direction of one electron in the
direction of the laser polarization ($\theta_1\!=\!0^\circ$) and
calculating the probability for the second electron to be emitted into the
forward half-space $\theta_2\leq\pi/2$ or backward half-space
$\theta_2>\pi/2$. The probabilities thus defined are
\begin{multline}\label{eq:Pplusminus}
  P^\pm(E_1,E_2) =\\
   4\pi^2 \int_{\genfrac{}{}{0pt}{}{\theta_2<\pi/2}{\theta_2>\pi/2}} P(E_1,E_2,\theta_1\!=\!0^\circ,\theta_2) \sin\theta_2 \dd\theta_2\eqcomma
\end{multline}
where the factor $4\pi^2$ stems from integration over $\phi_1$ and $\phi_2$.
The forward-backward asymmetry is then given by
\begin{equation}\label{eq:Pasym}
  \Pasym(E_1,E_2) = \frac{P^+(E_1,E_2)-P^-(E_1,E_2)}{P^+(E_1,E_2)+P^-(E_1,E_2)}\eqcomma
\end{equation}
which varies in the range $[-1,1]$. Values close to zero indicate vanishing correlation between 
the electrons, while large absolute values identify strong angular correlations.
Positive values ($\Pasym>0$) indicate a preference for ejection of both electrons in the 
same direction while negative values ($\Pasym<0$) indicate ejection in opposite directions. 
Note that $\Pasym(E_1,E_2)$ is not symmetric under exchange of $E_1$ and $E_2$, as the 
emission direction of the electron with energy $E_1$ is fixed in the laser 
polarization direction.
Analogously, the reduced one-electron asymmetry $\Pasym(E_1)$ can be determined
by integrating $P^\pm(E_1,E_2)$ over the energy of the second electron, \ie 
$P^\pm(E_1)=\int P^\pm(E_1,E_2) \dE_2$, and 
$\Pasym(E_1) = (P^+(E_1)-P^-(E_1))/(P^+(E_1)+P^-(E_1))$.

\begin{figure*}[tb]
  \centering
  \includegraphics[width=0.75\linewidth]{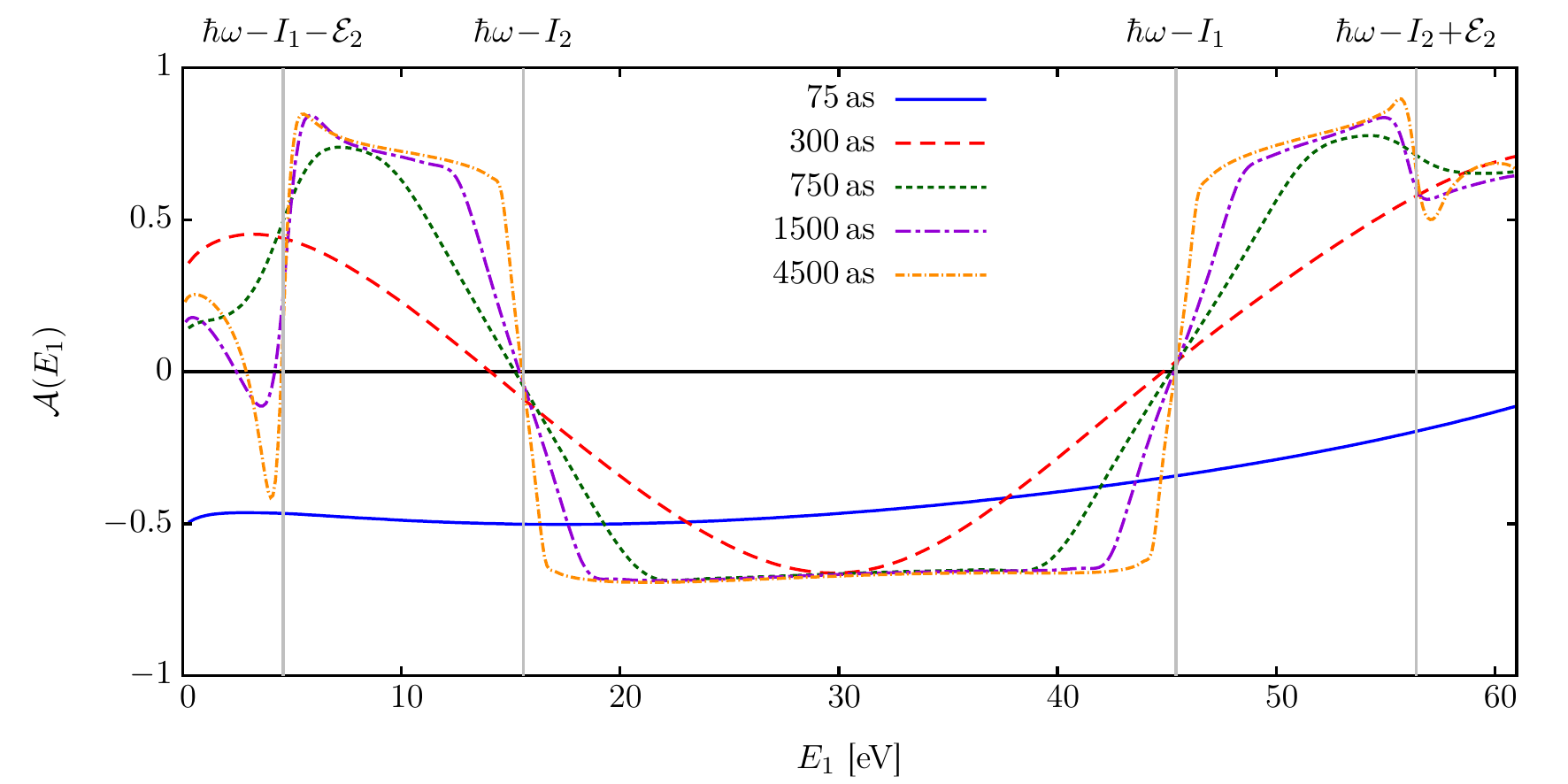}
  \caption{Forward-backward asymmetry $\Pasym(E_1)$ for TPDI by an XUV pulse at $\hw=70\ev$, 
  for different pulse durations $\Tp$. The gray lines show the expected positions of the peaks
  for the sequential process (with and without shake-up).}
  \label{fig:asym_dur_scan}
\end{figure*}

\autoref{fig:asym_dur_scan} shows the asymmetry of
TPDI at $\hw=70\ev$ photon energy for pulses of different duration
$\Tp$, from $\Tp=75\as$ up to $\Tp=4500\as$.  For the shortest
pulses, the electrons are dominantly ejected in opposite directions
independent of energy, as observed previously \cite{FeiNagPaz2008b}.
As the duration is increased, a stable pattern
emerges: at the ``sequential'' peaks, the electrons are essentially 
uncorrelated, leading to vanishing asymmetry. As most electrons are ejected
in this channel, the total (energy-integrated) asymmetry is very small for long pulses.
However, for energies in between the two main peaks at $E_1=\hw-I_1$ and
$E_2=\hw-I_2$, the electrons are emitted in opposite
directions. This is precisely because these final state energies 
are reached only when the two electrons are ejected in such a configuration.
This back-to-back Wannier-like emission near equal energy sharing remains pronounced 
even for long pulses.

\begin{figure*}[tb]
  \centering
  \includegraphics[width=0.75\linewidth]{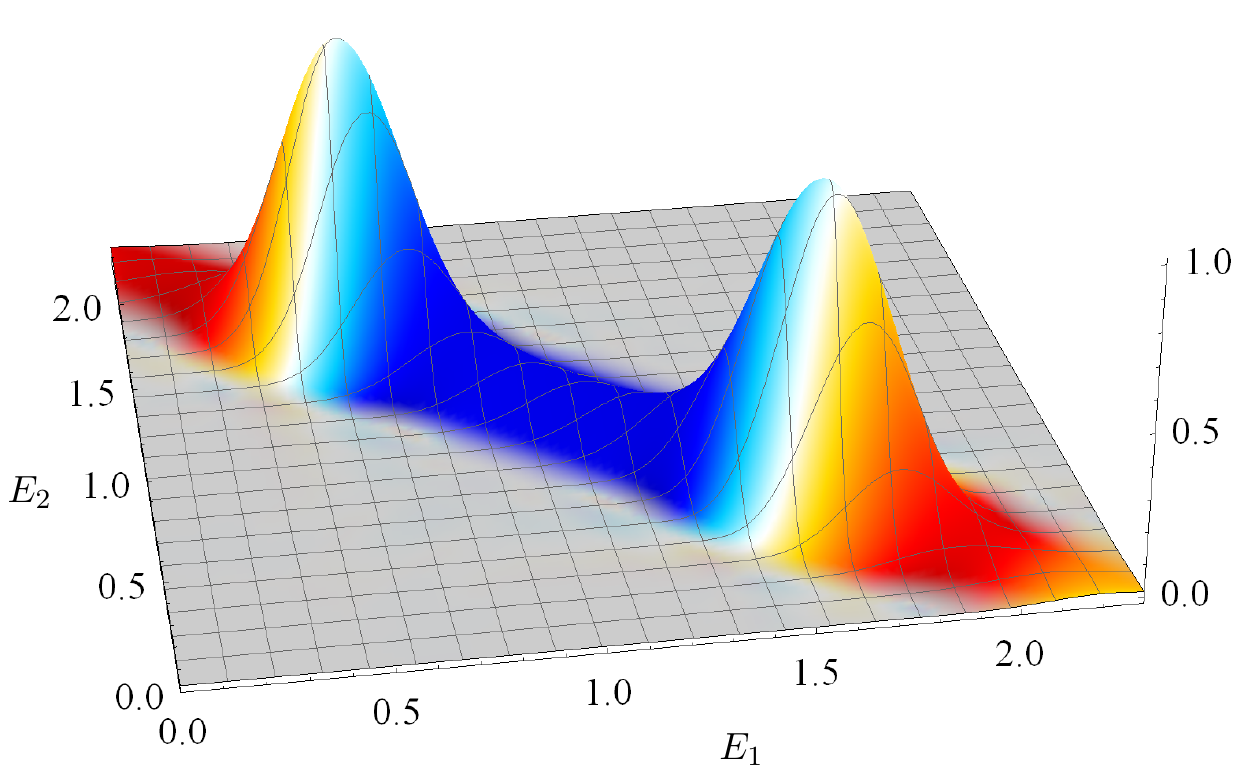}
  \caption{Combined double ionization probability $\PDI(E_1,E_2)$ and
  forward-backward asymmetry $\Pasym(E_1,E_2)$ after TPDI by an XUV pulse 
  at $\hw=70\ev$ with a duration of $450\as$. The $z$-axis gives $\PDI(E_1,E_2)$ 
  (in arbitrary units), while the color encodes the asymmetry, 
  with cyan to blue signifying negative values (ejection in opposite direction) and yellow to red 
  signifying positive values (ejection in the same direction). Vanishing $\Pasym$ corresponds to white.
  For energies where $\PDI(E_1,E_2)$ is negligible, the color is set to gray.}
  \label{fig:asym_2d}
\end{figure*}

For energies outside the energy interval delimited by the sequential peaks, the
asymmetry is equally strong, but now positive pointing to the same emission
direction for both electrons. When the second electron is emitted in
the same direction as the first one, the well-known post-collision interaction 
\cite{BarBer1966,GerMorNie1972,RusMeh1986,ArmTulAbe1987} 
tends to increase the asymmetric sharing of the
available energy \cite{FeiNagPaz2008b}. The dividing line between the two different regimes
of ejection in opposite or in the same direction is quite sharp and
lies directly at the position of the sequential peaks.
A more complete representation of the two-electron energy and angular 
correlations is presented in \autoref{fig:asym_2d} for a pulse duration
of $\Tp=450\as$. While the height gives the joint probability $\PDI(E_1,E_2)$, the
color represents the asymmetry distribution $\Pasym(E_1,E_2)$. The borderline
between positive and negative $\Pasym$ (\ie $\Pasym\approx0$, white)
is precisely near the peaks associated with the sequential process.
In the central region in between the ``sequential'' peaks the emission is
preferentially on opposite sides while emission into the same hemisphere
prevails outside the main peaks. For completeness we note that in the region 
between the two main peaks, only electrons emitted in opposite direction are observed 
both in ``sequential'' ($\hw>54.4\ev$) and ``nonsequential'' ($39.5\ev<\hw<54.4\ev$) 
TPDI \cite{FeiNagPaz2008}. The main difference is that 
in nonsequential TPDI, only that region is energetically accessible, such 
that no other angular configurations are observed.

%% file: shakeup.tex
\section{Shake-up interferences}\label{sec:shakeup}
We return now to the additional structures at higher
($E\approx\hw-I_2+\E_2$) and lower ($E\approx\hw-I_1-\E_2$)
energies visible in Figs.\ \ref{fig:pdi_dur_scan} and \ref{fig:asym_dur_scan}.
They correspond to shake-up satellites in $\Hep$ which can
serve as intermediate states in sequential TPDI.
In the shake-up process, the $\Hep$ ion is left in an
excited state, while the free electron obtains an energy
of $E_1'=\hw-I_1-\E_n$ (with $\E_n$ the excitation energy to the $n$-th
shell of $\Hep$). 
In the long-pulse limit, this simply leads to the appearance of shake-up 
satellite lines at energies $E_1'$ and
$E_2'=\hw-I_2+\E_n$ in the one-electron energy spectrum.
For ultrashort pulses, however, the nonsequential (or direct) double 
ionization channel becomes available as well and can lead to the same
final states. 
Post-collision interactions (PCI) lead to a broad distribution of
electron energies (see \autoref{sec:pci}), so that
the electrons can obtain the same final energies of $E_1^{PCI}=E_2'$ and $E_2^{PCI}=E_1'$ as 
the electrons emitted via $\Hep(nl)$ in the sequential process. 
Both indistinguishable pathways lead to the same final state and thus
to an interference pattern in the 
double ionization yield, as observed in \autoref{fig:pdi_dur_scan} and \autoref{fig:asym_dur_scan}.
This interference bears some resemblance to the well-known
exchange interference between \eg photo-electrons and 
Auger electrons \cite{VegMac1994,Rea1997,GouEckPet1993,RouRioAva2003}.
There is, however, a fundamental difference: while the exchange interference
is intrinsically controlled by atomic parameters, namely the energy
and lifetime (width) of the Auger electron, the novel interference
observed here is truly a dynamical effect present only for short pulses and 
can be controlled by the pulse duration $\Tp$.

\begin{figure*}[tbp]
  \centering
  \includegraphics[width=\linewidth]{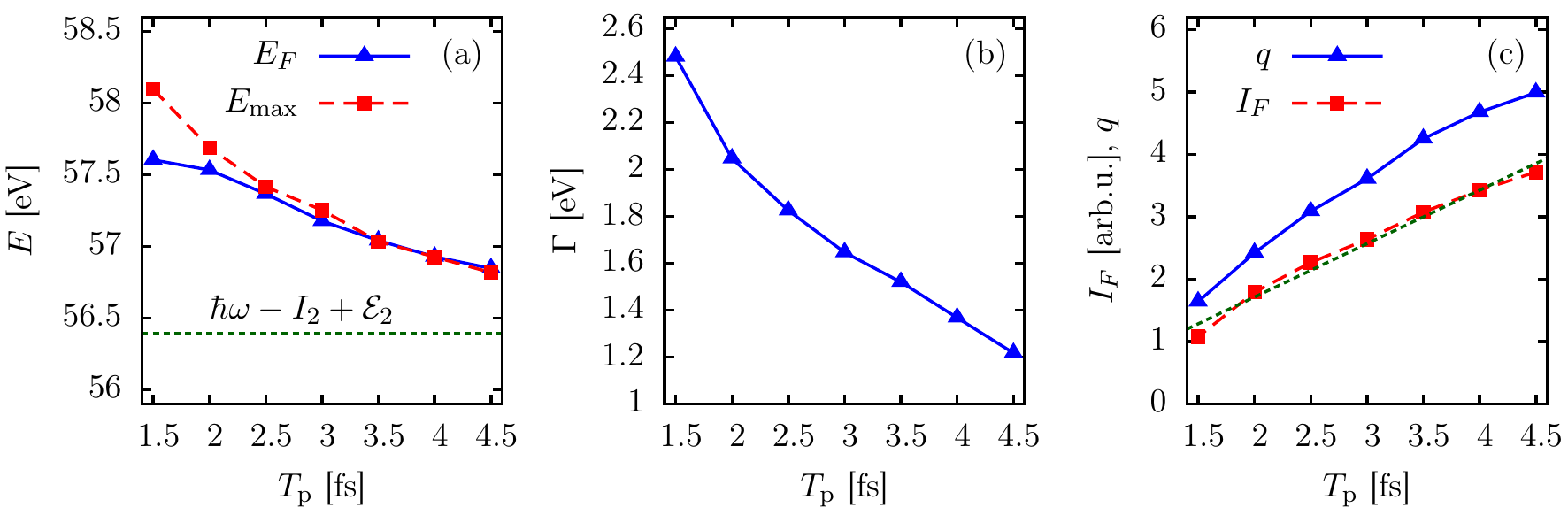}
  \caption{Parameters of the shake-up interference peaks around $57\ev$ for TPDI by an XUV pulse 
  at $\hw=70\ev$ obtained from fitting to a Fano lineshape.
  (a) Fano resonance energy $E_F$ and position $\Emax$ of the maximum in the spectrum,
  (b) width $\Gamma$, (c) Fano parameter $q$ and integrated yield $I_F$ from the shake-up pathway.
  See text for details.}
  \label{fig:shake_fanofits}
\end{figure*}

As the dependence of the yield on pulse duration is different for
the different channels (proportional to $\Tp$ for the nonsequential channel,
proportional to $\Tp^2$ in the sequential channel), the observed spectrum strongly
changes with pulse duration. For short pulses ($\Tp<500\as$, \cf\autoref{fig:pdi_dur_scan}),
the yield is completely dominated by the nonsequential channel without any trace of a shake-up interference.
As the pulse duration is increased, the sequential channel with shake-up becomes
increasingly important. As expected from the interference of a relatively sharp peak with a smooth
background, the peak resembles a Fano lineshape \cite{Fano61}. Thus, the position of the maximum is shifted from 
the position expected in the limit of infinitely long pulses.
Even for relatively long pulses ($\Tp=4.5\fs$), similar to those produced in free electron lasers, the position 
of the shake-up peak in the one-electron energy spectrum $\PDI(E)$ is shifted by 
a considerable fraction of an eV.
The structural similarity to a Fano resonance (a quasi-discrete resonance due to the shake-up 
intermediate state embedded in a smooth continuum due to the direct double ionization) 
suggests to characterize the interference in terms of Fano resonance parameters for the position $E_F(\Tp)$, 
width $\Gamma(\Tp)$, and asymmetry parameter $q(\Tp)$, as well as its strength $I_F(\Tp)$ (\autoref{fig:shake_fanofits}). 
To apply Fano's parametrization \cite{Fano61}, 
the calculated energy spectrum $\PDI(E)$ is divided by the nonresonant spectrum $\PDInonres(E)$, taken to be
proportional to the singly differential cross section as predicted from the model 
by Horner \etal \cite[\equationautorefname\ (8)]{HorMorRes2007}. Away from the peaks, this fits 
the form of the spectrum very well. 
A background contribution $c_{bg}$ is added to account for 
the different angular distributions of the different channels, which prevent complete interference.
This gives 
\begin{align}\label{eq:fano_fit}
\frac{\PDI(E)}{\PDInonres(E)} &\approx c_{bg} + c_F \frac{(q \Gamma/2 + E - E_F)^2}{(E-E_F)^2+(\Gamma/2)^2}\eqcomma
\end{align}
The simple fitting procedure used here only works well for pulse durations $\Tp\geq1.5\fs$, as 
for shorter pulses, the employed approximation for the ``nonresonant'' background breaks down,
and the shake-up peak itself is less strong and considerably broadened.
\autoref{fig:shake_fanofits} illustrates the dependence 
of the obtained parameters on the pulse duration, confirming the expected behavior: 
for long pulses, the peaks converge to the satellite lines, 
\ie Lorentzians of vanishing width, such that $E_F\to\hw-I_2+\E_n$ ($E_F\to\hw-I_1-\E_n$), $\Gamma\to0$, $\abs{q}\gg1$.
The overall strength $I_F$ of the shake-up peak relative to the nonresonant background is obtained from 
the integral over the Fano lineshape, $I_F \propto c_F (q^2-1)\Gamma$.
This behaves approximately linear with $\Tp$, confirming the scaling of the sequential shake-up 
channel with $\Tp^2$ versus the scaling of the nonresonant background with $\Tp$ (\autoref{fig:shake_fanofits}c).
Also shown in \autoref{fig:shake_fanofits}a is the position $\Emax$ of the maximum of the spectrum $\PDI(E)$
without any further processing. 

Such effects could possibly be observed in FEL pulses, 
which reach focused intensities of up to $10^{16}\Wcm$. 
To confirm that the results shown here (calculated for $10^{12}\Wcm$) also apply for these high
intensities, we performed an additional calculation at a peak intensity of $I_0=5\cdot 10^{15}\Wcm$ with a
pulse duration of $\Tp=4.5\fs$. The shape of the differential yield $\PDI(E)$ (not shown) is almost unchanged
compared to the result at $10^{12}\Wcm$ peak intensity, even though the ground state survival probability
is only $32\%$. The total double ionization probability is $\PDI=36\%$, \ie more than a third of
the helium atoms in the laser focus are doubly ionized. Even though the yield in the shake-up 
peak is only $0.6\%$ of the total yield for that duration, this could be seen in experiment
as only the integrated one-electron energy spectrum has to be observed.
Moreover, from the position, strength and asymmetry of the interference peaks, information
on the poorly known pulse duration of FEL pulse ``bursts'' could possibly be deduced.

\begin{figure*}[tb]
  \centering
  \includegraphics[width=\linewidth]{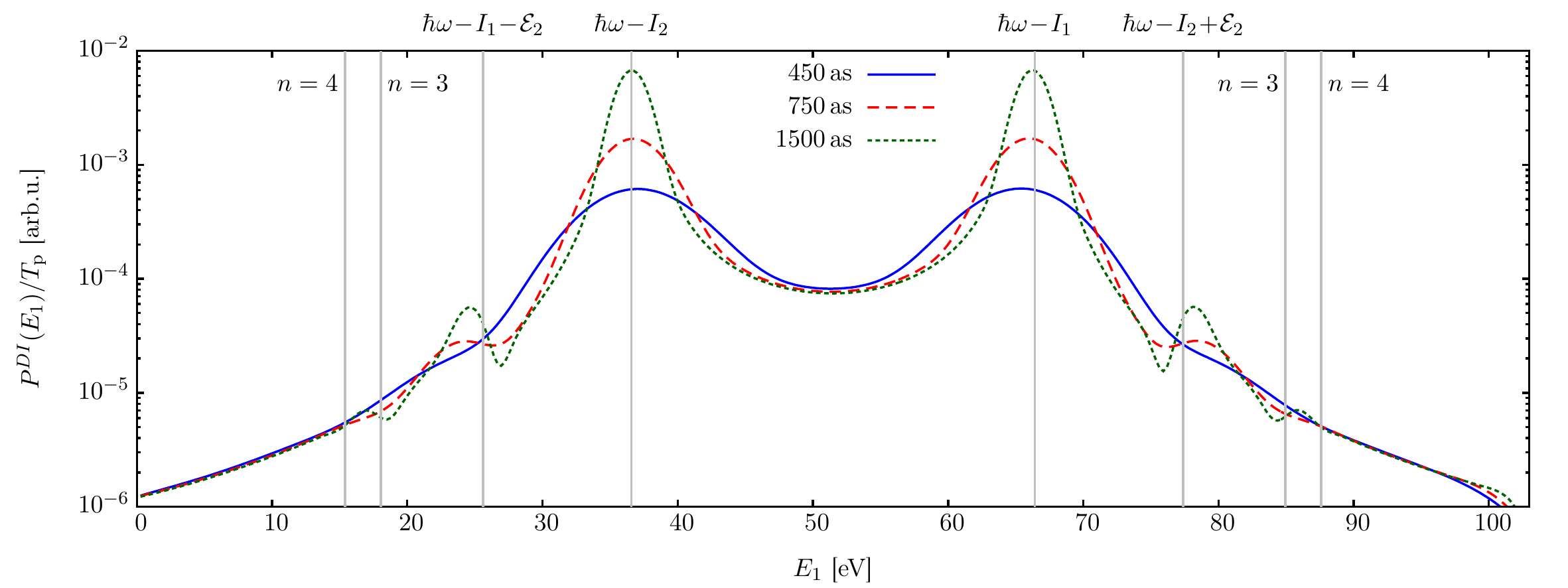}
  \caption{Double ionization (DI) rate $\PDI(E)/\Tp$ (\ie DI probability divided by the pulse duration) 
  for TPDI by an XUV pulse at $\hw=91\ev$ with different pulse durations $\Tp$. Shake-up peaks up to
  $n=3$ are visible.}
  \label{fig:91ev_pdi_dur_scan}
\end{figure*}

The results shown up to now were obtained at a photon energy of $\hw=70\ev$, 
where only the $n=2$ shake-up channel plays a role.
While the qualitative behaviour of each shake-up peak is expected to be independent of $\hw$,
new intermediate ionic states $\ket{nl}$ become accessible at $\hw>I_1+\E_n$,
converging to $\hw>-E_0$ for $n\to\infty$ (where $E_0\!\approx\!-79\ev$ is 
the ground state energy of helium).
This is demonstrated in \autoref{fig:91ev_pdi_dur_scan} at a photon energy of $\hw=91\ev$.
As the shake-up probability strongly decreases with increasing $n$, only the peaks associated
with $n=2$ and $n=3$ can clearly be identified at the pulse lengths used here (up to $\Tp=1.5\fs$).
For longer pulses, more highly excited states would start to play a role as well.  
In that case, one would need to take into account that the peaks for higher $n$ overlap with
each other as well as with the nonresonant background.

It should be noted that in order to observe these interference effects, 
the asymptotic vectorial momenta $\kone,\ktwo$ (\ie not only the 
asymptotic energies $(E_1,E_2)$) of the two pathways have to coincide. 
The shake-up channel has an angular distribution considerably
different from that of nonsequential channel, such that only
partial interference between the final states is expected. This leads to a rich structure
in the observed angular distributions (not shown), a more detailed analysis of which is in progress.
During the preparation of this contribution, we became aware of work by Palacios \etal \cite{PalResMcc2009}
who also observe the interference between these different channels.

%% file: summary.tex
\section{Summary}\label{sec:summary}
We have presented a detailed study of the dynamics of the two-photon double
ionization process in helium in the so-called ``sequential'' energy regime for a wide range
of ultrashort pulse durations ($75\as$ to $4.5\fs$). We have shown how electron interaction and 
thereby correlation enforced by the short pulse duration influences the observed 
energy spectra and angular distributions.

The one-electron ionization rate $\PDI(E)/\Tp$ converges to a 
stable value with increasing pulse duration for energies
away from the sequential peaks ($E=\hw-I_1$ and $E=\hw-I_2$), giving
rise to a well-defined (direct) differential double ionization
cross section. However, near the peaks where the sequential process
is allowed, $\PDI(E)/\Tp$ grows with $\Tp$. We have thus observed
a non-uniform scaling of the double ionization probability
with $\Tp$. Even though in this spectral range the sequential
process is allowed, both the direct and sequential co-exist, giving
rise to interferences which are induced by the short time correlation
between the two emission events.
The nonsequential channel without shake-up and the sequential shake-up channel, where the intermediate
state after one-photon absorption is an excited state of the $\Hep$ ion, can interfere. In attosecond pulses, only
the nonsequential channel contributes, while in long pulses (longer than the $4.5\fs$ used here), 
the sequential shake-up channel would dominate. For pulse durations of a few femtoseconds,
as obtained in free electron lasers, the two channels are similarly important, such that
interference can be clearly observed. 
This interferences may open up the possibility to measure the duration of ultrashort XUV pulses
in the femtosecond regime. 

We have also found that the angular distributions in the final states populated
by nonsequential processes are strongly correlated. In ultrashort pulses, where 
the TPDI process is necessarily nonsequential, the favored emission channel is 
the Wannier ridge riding mode of back-to-back emission at equal energies (\cf\cite{FeiNagPaz2008b}).
In longer pulses, back-to-back emission is strongly favored in the region close 
to equal energy sharing, while for strongly asymmetric energy sharing, the electrons are
primarily emitted in the same direction.

\ack
J.F., S.N., R.P., E.P., and J.B.\ acknowledge support by the FWF-Austria, Grant No.\ SFB016.
Computational time provided under Institutional Computing at Los Alamos.
The Los Alamos National Laboratory is operated by Los Alamos National Security, LLC 
for the National Nuclear Security Administration of the U.S.\ Department of Energy under Contract No.~DE-AC52-06NA25396.

%% file: tpdi_long.bbl
\begin{thebibliography}{10}

\bibitem{FLASH2007}
Ackermann W \emph{et~al.} 2007 \emph{Nat. Photonics} \textbf{1} 336

\bibitem{DroZepGop2006}
Dromey B \emph{et~al.} 2006 \emph{Nat. Phys.} \textbf{2} 456

\bibitem{GouSchHof2008}
Goulielmakis E \emph{et~al.} 2008 \emph{Science} \textbf{320} 1614

\bibitem{HenKieSpi2001}
Hentschel M, Kienberger R, Spielmann C, Reider G~A, Milosevic N, Brabec T,
  Corkum P, Heinzmann U, Drescher M and Krausz F 2001 \emph{Nature}
  \textbf{414} 509

\bibitem{NabHasTak2005}
Nabekawa Y, Hasegawa H, Takahashi E~J and Midorikawa K 2005 \emph{Phys. Rev.
  Lett.} \textbf{94} 043001

\bibitem{NauNeeSok2004}
Naumova N~M, Nees J~A, Sokolov I~V, Hou B and Mourou G~A 2004 \emph{Phys. Rev.
  Lett.} \textbf{92} 063902

\bibitem{NomHorTza2008}
Nomura Y \emph{et~al.} 2008 \emph{Nat. Phys.} \textbf{advanced online
  publication} doi:10.1038/nphys1155

\bibitem{SanBenCal2006}
Sansone G \emph{et~al.} 2006 \emph{Science} \textbf{314} 443

\bibitem{SerYakSer2007}
Seres J, Yakovlev V~S, Seres E, Streli C, Wobrauschek P, Spielmann C and Krausz
  F 2007 \emph{Nat. Phys.} \textbf{3} 878

\bibitem{ZhaLytPop2007}
Zhang X, Lytle A~L, Popmintchev T, Zhou X, Kapteyn H~C, Murnane M~M and Cohen O
  2007 \emph{Nat. Phys.} \textbf{3} 270

\bibitem{FeiNagPaz2008b}
Feist J, Nagele S, Pazourek R, Persson E, Schneider B~I, Collins L~A and
  Burgd\"{o}rfer J 2008 \emph{arXiv.org:physics} 0812.0373

\bibitem{ParSmyTay1998}
Parker J~S, Smyth E~S and Taylor K~T 1998 \emph{J. Phys. B} \textbf{31} L571

\bibitem{BraDoeCoc1998}
Br\"{a}uning H \emph{et~al.} 1998 \emph{J. Phys. B} \textbf{31} 5149

\bibitem{BriSch2000}
Briggs J~S and Schmidt V 2000 \emph{J. Phys. B} \textbf{33} R1

\bibitem{ByrJoa1967}
Byron F~W and Joachain C~J 1967 \emph{Phys. Rev.} \textbf{164} 1

\bibitem{MalSelKaz2000}
Malegat L, Selles P and Kazansky A~K 2000 \emph{Phys. Rev. Lett.} \textbf{85}
  4450

\bibitem{ProSha1993}
Proulx D and Shakeshaft R 1993 \emph{Phys. Rev. A} \textbf{48} R875

\bibitem{LeiGroEng2000a}
Lein M, Gross E~K~U and Engel V 2000 \emph{Phys. Rev. Lett.} \textbf{85} 4707

\bibitem{RudJesErg2007}
Rudenko A, de~Jesus V~L~B, Ergler T, Zrost K, Feuerstein B, Schr\"{o}ter C~D,
  Moshammer R and Ullrich J 2007 \emph{Phys. Rev. Lett.} \textbf{99} 263003

\bibitem{StaRuiSch2007}
Staudte A \emph{et~al.} 2007 \emph{Phys. Rev. Lett.} \textbf{99} 263002

\bibitem{AntFouPir2008}
Antoine P, Foumouo E, Piraux B, Shimizu T, Hasegawa H, Nabekawa Y and
  Midorikawa K 2008 \emph{Phys. Rev. A} \textbf{78} 023415

\bibitem{ColPin2002}
Colgan J and Pindzola M~S 2002 \emph{Phys. Rev. Lett.} \textbf{88} 173002

\bibitem{FeiNagPaz2008}
Feist J, Nagele S, Pazourek R, Persson E, Schneider B~I, Collins L~A and
  Burgd\"{o}rfer J 2008 \emph{Phys. Rev. A} \textbf{77} 043420

\bibitem{FenHar2003}
Feng L and van~der Hart H~W 2003 \emph{J. Phys. B} \textbf{36} L1

\bibitem{FouLagEdaPir2006}
Foumouo E, Lagmago~Kamta G, Edah G and Piraux B 2006 \emph{Phys. Rev. A}
  \textbf{74} 063409

\bibitem{GuaBarSch2008}
Guan X, Bartschat K and Schneider B~I 2008 \emph{Phys. Rev. A} \textbf{77}
  043421

\bibitem{HasTakNabIsh2005}
Hasegawa H, Takahashi E~J, Nabekawa Y, Ishikawa K~L and Midorikawa K 2005
  \emph{Phys. Rev. A} \textbf{71} 023407

\bibitem{HorMccRes2008}
Horner D~A, McCurdy C~W and Rescigno T~N 2008 \emph{Phys. Rev. A} \textbf{78}
  043416

\bibitem{HuCoCo2005}
Hu S~X, Colgan J and Collins L~A 2005 \emph{J. Phys. B} \textbf{38} L35

\bibitem{IvaKhe2007}
Ivanov I~A and Kheifets A~S 2007 \emph{Phys. Rev. A} \textbf{75} 033411

\bibitem{NikLam2001}
Nikolopoulos L~A~A and Lambropoulos P 2001 \emph{J. Phys. B} \textbf{34} 545

\bibitem{NikLam2007}
Nikolopoulos L~A~A and Lambropoulos P 2007 \emph{J. Phys. B} \textbf{40} 1347

\bibitem{ProManMar2007}
Pronin E~A, Manakov N~L, Marmo S~I and Starace A~F 2007 \emph{J. Phys. B}
  \textbf{40} 3115

\bibitem{SorWelBob2007}
Sorokin A~A, Wellhofer M, Bobashev S~V, Tiedtke K and Richter M 2007
  \emph{Phys. Rev. A} \textbf{75} 051402(R)

\bibitem{BarWanBur2006}
Barna I~F, Wang J and Burgd\"{o}rfer J 2006 \emph{Phys. Rev. A} \textbf{73}
  023402

\bibitem{FouAntBac2008}
Foumouo E, Antoine P, Bachau H and Piraux B 2008 \emph{New J. Phys.}
  \textbf{10} 025017

\bibitem{IshMid2005}
Ishikawa K~L and Midorikawa K 2005 \emph{Phys. Rev. A} \textbf{72} 013407

\bibitem{LauBac2003}
Laulan S and Bachau H 2003 \emph{Phys. Rev. A} \textbf{68} 013409

\bibitem{PirBauLau2003}
Piraux B, Bauer J, Laulan S and Bachau H 2003 \emph{Eur. Phys. J. D}
  \textbf{26} 7

\bibitem{HorMorRes2007}
Horner D~A, Morales F, Rescigno T~N, Mart\'{i}n F and McCurdy C~W 2007
  \emph{Phys. Rev. A} \textbf{76} 030701(R)

\bibitem{UllMosDor2003}
Ullrich J, Moshammer R, Dorn A, D\"{o}rner R, Schmidt L~P~H and
  Schmidt-B\"{o}cking H 2003 \emph{Rep. Prog. Phys.} \textbf{66} 1463

\bibitem{PinRobLoc2007}
Pindzola M~S \emph{et~al.} 2007 \emph{J. Phys. B} \textbf{40} R39

\bibitem{MccHorRes2001}
McCurdy C~W, Horner D~A and Rescigno T~N 2001 \emph{Phys. Rev. A} \textbf{63}
  022711

\bibitem{ResMcc2000}
Rescigno T~N and McCurdy C~W 2000 \emph{Phys. Rev. A} \textbf{62} 032706

\bibitem{Schneider05}
Schneider B~I and Collins L~A 2005 \emph{J. Non-Cryst. Solids} \textbf{351}
  1551

\bibitem{SchColHu2006}
Schneider B~I, Collins L~A and Hu S~X 2006 \emph{Phys. Rev. E} \textbf{73}
  036708

\bibitem{Lefo90}
Leforestier C \emph{et~al.} 1991 \emph{J. Comp. Phys.} \textbf{94} 59

\bibitem{ParkLight86}
Park T~J and Light J~C 1986 \emph{J. Chem. Phys.} \textbf{85} 5870

\bibitem{SmyParTay1998}
Smyth E~S, Parker J~S and Taylor K~T 1998 \emph{Comput. Phys. Commun.}
  \textbf{114} 1

\bibitem{BarBer1966}
Barker R~B and Berry H~W 1966 \emph{Phys. Rev.} \textbf{151} 14

\bibitem{GerMorNie1972}
Gerber G, Morgenstern R and Niehaus A 1972 \emph{J. Phys. B} \textbf{5} 1396

\bibitem{RusMeh1986}
Russek A and Mehlhorn W 1986 \emph{J. Phys. B} \textbf{19} 911

\bibitem{ArmTulAbe1987}
Armen G~B, Tulkki J, \r{A}berg T and Crasemann B 1987 \emph{Phys. Rev. A}
  \textbf{36} 5606

\bibitem{GouEckPet1993}
de~Gouw J~A, van Eck J, Peters A~C, van~der Weg J and Heideman H~G~M 1993
  \emph{Phys. Rev. Lett.} \textbf{71} 2875

\bibitem{Rea1997}
Read F~H 1977 \emph{J. Phys. B} \textbf{10} L207

\bibitem{RouRioAva2003}
Rouvellou B, Rioual S, Avaldi L, Camilloni R, Stefani G and Turri G 2003
  \emph{Phys. Rev. A} \textbf{67} 012706

\bibitem{VegMac1994}
V\'{e}gh L and Macek J~H 1994 \emph{Phys. Rev. A} \textbf{50} 4031

\bibitem{Fano61}
Fano U 1961 \emph{Phys. Rev.} \textbf{124} 1866

\bibitem{PalResMcc2009}
Palacios A, Rescigno T~N and McCurdy C~W \emph{(unpublished)}

\end{thebibliography}
